\documentclass[aps,pra,twocolumn,superscriptaddress,floatfix,showkeys
]{revtex4-2}

\usepackage{graphicx}
\usepackage{dcolumn}
\usepackage{bm}
\usepackage[usenames,dvipsnames]{xcolor}
\usepackage{amsmath}
\usepackage{physics}
\usepackage{makecell}
\usepackage{xcolor}
\usepackage{colortbl}

\begin{document}


\title{\textbf{
Analytical model for polarization transfer during gas-phase collision events in spin-exchange optical pumping: Spin-$\frac{1}{2}$ $^{129}$Xe versus spin-$\frac{3}{2}$ $^{131}$Xe
} 
}%

\author{Perttu Hilla}
\email{perttu.hilla@oulu.fi}
\author{Rajgowrav Cheenikundil}%
\email{rajgowrav.cheenikundil@oulu.fi}
\author{Juha Vaara}%
\email{juha.vaara@oulu.fi}
\affiliation{NMR Research Unit, P.O.~Box 3000, FI-90014 University of Oulu, Finland
}%



\date{\today}

\begin{abstract}


Spin-exchange optical pumping (SEOP) is a method 
for producing spin-hyperpolarized noble gas nuclei, such as $^{129}$Xe and $^{131}$Xe, which are used in various magnetic resonance
applications from fundamental physics to quantum sensing and medical imaging. In SEOP, optically polarized alkali-metal atoms transfer their spin polarization to the noble gas nuclei in gas-phase collision events via the hyperfine coupling (HFC) between the alkali valence electron and the noble gas nucleus. 
While the polarization transfer physics of
spin $I = \frac{1}{2}$ nuclei, such as $^{129}$Xe, is relatively well understood, that of spin $I > \frac{1}{2}$ nuclei, such as $^{131}$Xe 
($I = \frac{3}{2}$), has been far less studied, and no rigorous theoretical model has been presented to date.
To this end, we derive a simple analytical model for the
upper limit, neglecting relaxation, of the
SEOP polarization transfer, 
applicable to noble gases with arbitrary nuclear spin.
Analytical evaluation of the Baker-Campbell-Hausdorff expansion 
for the time evolution of the spin density operator $\hat{\rho}(t)$
reveals that only even-order terms 
in the HFC
contribute to the polarization transfer, with the leading-order quadratic term being the most significant.
We obtain a result similar to that derived for the spin-exchange cross section by Herman [Phys.\ Rev.\ {\bf 137}, A 1062 (1965)], 
but in a more general framework for the time evolution of $\hat{\rho}(t)$ 
that is also more familiar to magnetic resonance researchers.
The model is applied to understand the difference in the polarization transfer efficiency between $^{129}$Xe and $^{131}$Xe, 
yielding results in
agreement with previous experiments.
We also validate the model by comparison to detailed numerical
multiscale simulations of the SEOP process, where full quantum-chemically computed spin Hamiltonians sampled from molecular 
dynamics simulations of the gas-phase
collision events are used to propagate the spin dynamics.
\end{abstract}

\maketitle


\section{Introduction}



Spin-exchange optical pumping (SEOP)~\cite{wal97} is a means of producing highly spin-polarized samples of noble gas nuclei, useful for applications in magnetometry~\cite{bud07}, nuclear magnetic resonance spectroscopy~\cite{mee15}, magnetic resonance imaging~\cite{goo02}, and fundamental physics~\cite{bro17, 2025_Walter_PhysRevD_Search_for_Axionlike_Dark_Matter_Using_Liquid-State_Nuclear_Magnetic_Resonance}. 
In SEOP,
electronic spin polarization is induced on alkali-metal vapor by irradiation with circularly polarized light, and the polarization is transferred to the nuclei of noble gas atoms in gas-phase interaction events, 
either
short-lifetime binary collisions~\cite{cat92} (which can also be termed \emph{scattering} events)
or long-lived van der Waals (vdW) 
complexes~\cite{app98}, 
between the alkali metal and noble gas atoms. The mechanism of the transfer is hyperfine coupling (HFC) of the 
unpaired electron spin of the 
alkali metal atom
to the noble gas nucleus. 

In many cases the rubidium-85 or -87 isotopes are used as the alkali metal, and the target noble gas nuclei are most often the spin-$\frac{1}{2}$ isotopes $^3$He~\cite{gen17} and $^{129}$Xe~\cite{kel21}. Particularly the latter isotope 
has gained
popularity due to its high natural abundance and availability, as well as the fact that the large, polarizable electron cloud of the xenon atom enables delicate response to various interactions of relevance in the applications. In contrast, there are much fewer SEOP studies of the quadrupolar, 
spin $I > \frac{1}{2}$
isotopes, such as the spin-$\frac{3}{2}$
$^{131}$Xe~\cite{vol80,kwo81,wu87,wu90,but94,mee98,don09,stu10,stu11,bul13,ver18,pet19,tra19,mol23,Lu2024}. 
These provide
a complementary view to the surroundings of the atom
due to the quadrupole interaction between the electric field gradient at the nuclear site and the electric quadrupole moment of the nucleus, {\em e.g.}, in lung imaging~\cite{lil15}
and for probing void spaces in porous materials~\cite{2001_Moudrakovski_JACS_131Xe_a_New_NMR_Probe_of_Void_Space_in_Solids}.
Further interest in $^{131}$Xe is found due to its  neutron scattering properties differing from $^{129}$Xe~\cite{Lu2024}, within 
$^{129}$Xe-$^{131}$Xe co-magnetometry~\cite{tra19}, and spin gyroscopes~\cite{wal16}.

It has been found experimentally that, in comparable conditions, $^{131}$Xe gains in SEOP a less strong polarization than $^{129}$Xe ~\cite{vol80,kwo81,but94,stu11,mol23,Lu2024}. The experimentally observed polarization transfer rate in SEOP
is 
affected by the spin relaxation of the Xe isotopes, which for the quadrupolar $^{131}$Xe is much faster than
for $^{129}$Xe~\cite{saa15,wu21,mol23}. The two isotopes differ, however, already in the actual polarization transfer during the interaction events, before relaxation takes place. One thinkable reason for the difference 
is in the smaller magnitude of the nuclear gyromagnetic ratio $\gamma$
of $^{131}$Xe
than that of $^{129}$Xe.
The HFC is directly proportional to $\gamma$, which 
could make one
{\em a priori\/} 
expect that the polarization 
transfer to
$^{131}$Xe would be 
proportionally
smaller than that 
to $^{129}$Xe by 
the ratio $|\gamma_{131}/\gamma_{129}|.$
The number of 
spin states is also
doubled in 
the spin-$\frac{3}{2}$ 
$^{131}$Xe as compared to 
the spin-$\frac{1}{2}$
$^{129}$Xe, which means that the detailed picture of the spin transitions~\cite{hil26} caused by the HFC in the interaction events with the alkali-metal atoms, differs for the two isotopes.
A further difference is that $^{131}$Xe is the only stable noble gas isotope with a positive nuclear gyromagnetic ratio~\cite{stu11}. 

In this paper we present a theoretical model for the 
SEOP polarization transfer 
during gas-phase interaction events,
valid for arbitrary nuclear spin $I$ of the noble gas,
as well as for both binary collisions and vdW complexes alike,
and apply it to compare the hyperpolarization processes of $^{129}$Xe and $^{131}$Xe.
The theoretical result implies that, in comparable experimental conditions, a smaller polarization 
transfer
takes place in the case of $^{131}$Xe than for $^{129}$Xe, and that the ratio between the two is dictated, on the one hand, by the spin quantum numbers and, on the other hand, the relative magnitude of the HFCs of the two isotopes. The latter is, then, very accurately determined 
by
the gyromagnetic ratios of $^{131}$Xe and $^{129}$Xe. 
Because relaxation always reduces the polarization gain, the theoretical result obtained here gives an \textit{upper limit} for the polarization transfer in SEOP.

Earlier we have introduced a multiscale simulation 
method for investigating the microscopic physics of the polarization transfer in SEOP~\cite{ran20,hil25,hil26}. The method combines molecular dynamics (MD) simulation of the gas mixture, quantum chemical (QC) electronic structure determination of the instantaneous spin Hamiltonians, and employing the time series of the Hamiltonians to propagate the spin dynamics (SD) of the alkali metal atom and the noble gas nucleus, in the interaction events between the two species. 
Here we employ this framework to test and verify the present theoretical prediction concerning the difference of the polarization transfer process between the two xenon isotopes,
in simplified SEOP conditions corresponding to those in Ref.~\cite{hil26}.
Finally, the findings are related to the available experimental data.  

The framework presented here provides a theoretical foundation for future SEOP studies of spin $I > \frac{1}{2}$ nuclei, offering new insight for the SEOP and magnetic resonance communities alike.

\section{Model for polarization transfer during gas-phase collision events}
\label{sec-theory}
\begin{figure*}[t]
    \centering
    \includegraphics[width=0.95\textwidth]{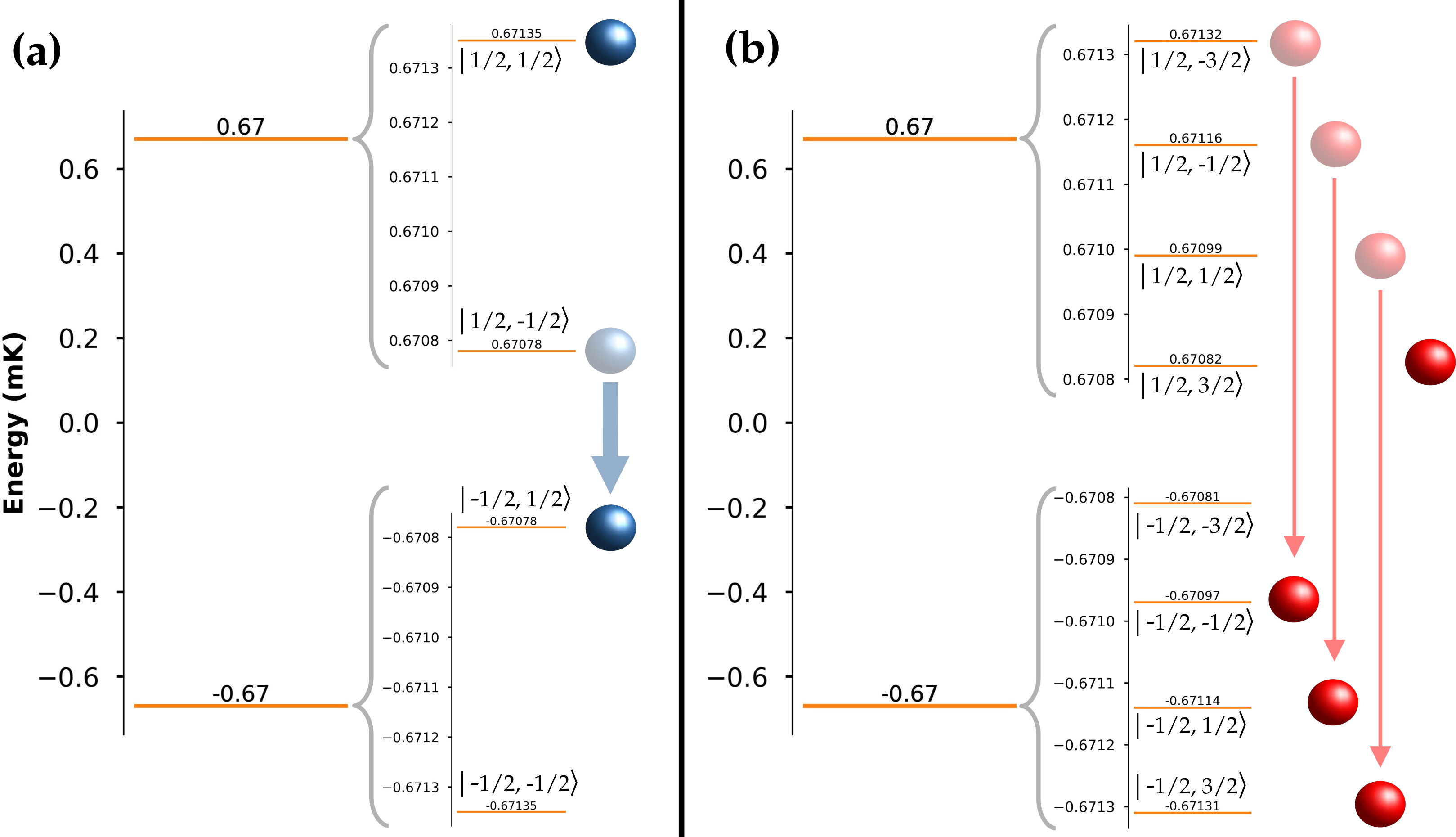}
    \caption{Energy level diagrams of the (a) electron-$^{129}$Xe  and (b) electron-$^{131}$Xe 
    spin systems with the spin states denoted as $\ket{m_S, m_K}$,
    using the energy eigenvalues of the spin
    Hamiltonian including the electron and nuclear Zeeman interactions
    at a magnetic field of $B=1$~mT.
    The hyperfine-allowed $\hat{S}_\pm \hat{K}_\mp$ flip-flop
    transitions are shown as arrows between the initial states implied by Eq.~(\ref{eq:initial_state})
    (transparent spheres) 
    and the final states (solid spheres).
    The thickness of the arrows illustrates the relative magnitudes of the HFCs of the two isotopes.
    }
    \label{fig:energy_diagram}
\end{figure*}

The time evolution of a spin system,
described by the spin density operator $\hat{\rho}(t)$,
is governed by the Liouville-von Neumann equation (LvNE)
\begin{equation}
    \frac{d}{dt} \hat{\rho}(t) = -i[\hat{H}(t), \hat{\rho}(t)],
\end{equation}
where
$\hat{H}(t)$ is the spin Hamiltonian of the system in units of $\hbar$.
The HFC between the Rb electron and the Xe nucleus
(or any other alkali electron and noble gas pair),
which
drives the polarization transfer in SEOP, is described by the Hamiltonian
\begin{equation}
    \hat{H}_{\rm HFC}^{(K)}(t) = \hat{\mathbf{S}} \cdot \mathbf{A}_K(t) \cdot \hat{\mathbf{K}},
\end{equation}
where 
$\hat{\mathbf{S}}$ and $\hat{\mathbf{K}}$ are the dimensionless
electron and nuclear spin operators, respectively,
and the spin quantum numbers $K_{129} =\frac{1}{2}$ and $K_{131}=\frac{3}{2}$ 
are used to label the $^{129}$Xe and $^{131}$Xe isotopes,
respectively.
$\mathbf{A}_K\big[R(t)\big] \equiv \mathbf{A}_K(t)$ is the 
isotope-specific
HFC tensor (in rad/s), which
depends on the Xe-Rb interatomic distance $R(t)$ during 
a collision event \cite{wal97,ran20,hil26}.
For theoretical analysis, the HFC can be treated as isotropic~\cite{hil26,wal98} with
a constant magnitude $A_K$ 
given by the average $A_K=\frac{1}{3}\langle{\rm Tr\mathbf{A}_K}\rangle$ over the range of distances occurring during the collision event,
such that
\begin{equation}
    \hat{H}_{\rm HFC}^{(K)} \approx A_K \hat{\mathbf{S}} \cdot \hat{\mathbf{K}}
     = A_K \bigg[ \hat{S}_z \hat{K}_z + \frac{1}{2} \bigl( \hat{S}_+ \hat{K}_- + \hat{S}_- \hat{K}_+ \bigr) \bigg].
\end{equation}
The relevant spin transitions are illustrated 
in the energy level diagrams
for the two Xe isotopes in Fig.~\ref{fig:energy_diagram}. 
Polarization transfer in $^{131}$Xe features three transitions as opposed to just one in $^{129}$Xe, in the presented two-spin model involving the unpaired electron and Xe nuclear spins but no Rb nuclear spin~\cite{hil26}. 

Since the HFC is the dominant interaction in a collision event, 
the LvNE describing the polarization transfer 
in SEOP
can be approximately written as 
\begin{equation}
    \frac{d}{dt} \hat{\rho}(t) \approx -i[\hat{H}_{\rm HFC}^{(K)}, \hat{\rho}(t)].
\end{equation}
Given an initial spin
state $\hat{\rho}(0)$, the time evolution of the system over a finite 
collision-event lifetime $\tau$
is given by the Baker-Campbell-Hausdorff (BCH) expansion
\begin{eqnarray}
\label{eq:bch_expansion}
    \hat{\rho}(\tau) & = &e^{-i\hat{H}_{\rm HFC}^{(K)}\tau}\hat{\rho}(0)e^{i\hat{H}_{\rm HFC}^{(K)}\tau} = \hat{\rho}(0) - i\tau [\hat{H}_{\rm HFC}^{(K)}, \hat{\rho}(0)] \nonumber \\
    &&\mbox{}+ \frac{(-i\tau)^2}{2!} [\hat{H}_{\rm HFC}^{(K)}, [\hat{H}_{\rm HFC}^{(K)}, \hat{\rho}(0)]] + \cdots.
\end{eqnarray}

In SEOP, the initial state of the system is 
to a good approximation
characterized by a fully polarized electron spin
in, depending on the polarization beam having either the $\sigma^+$ or $\sigma^-$ helicity~\cite{tie19}, the $|+\frac{1}{2}\rangle$ or $|-\frac{1}{2}\rangle$  angular momentum (Zeeman)
eigenstate 
and the Xe nuclear spin in a thermal equilibrium state. In reality, the degree of electron spin polarization is somewhat less than unity in SEOP experiments, and the populations of the states can be expected to follow the spin temperature
approximation~\cite{hap84}.
In a typical SEOP setup the magnetic field is of the order of a few mT, 
such that the thermal polarization of 
the noble gas can be safely neglected.
Hence, the initial state of Xe is well-approximated by a state of zero polarization.
In terms of the density operators, the initial states of the two spins are given by
\[
\hat{\rho}_S(0) = \frac{1}{2S + 1} \hat{1}_S \pm \hat{S}_z, \quad \hat{\rho}_K(0) = \frac{1}{2K + 1} \hat{1}_K,
\]
where $\hat{1}_S$ and $\hat{1}_K$ are the identity operators of dimensions $2S + 1$ and $2K + 1$, respectively.
Hence, the initial state of the system becomes
\begin{eqnarray}
    \label{eq:initial_state}
    \hat{\rho}(0) & = & \hat{\rho}_S(0) \otimes \hat{\rho}_K(0) 
    = \bigg( \frac{1}{2S + 1} \hat{1}_S \pm \hat{S}_z \bigg) \otimes \frac{1}{2K + 1} \hat{1}_K \nonumber \\
    & \Rightarrow & \hat{\rho}(0) = \pm \hat{S}_z \otimes \frac{1}{2K + 1} \hat{1}_K,
\end{eqnarray}
where we have used the fact that the identity operator $\hat{1}_S \otimes \hat{1}_K$ 
commutes with any other operator in the 
Liouville space \cite{Jeener_1982_AdvMagnOptReson_Superoperators_in_Magnetic_Resonance, Gyamfi_EurPhysJ_2020_Fundamentals_of_Quantum_Mechanics_in_Liouville_Space}
spanned by the two spins in question, and thus does not
contribute to the time evolution of the system.

The spin polarization level of the Xe nucleus at time $\tau$ is given by 
\begin{equation}
    \label{eq:xe_polarisation}
    p_K(\tau) \equiv {\rm Tr}\!\left[{\hat{p}_K \hat{\rho}(\tau)}\right],
\end{equation}
where $\hat{p}_K = \hat{K}_z / K$ is the $z$-polarization level
operator
\cite{2024_Bengs_JMagnRes_Hyperpolarisation_criteria_in_magnetic_resonance, 
2025_Barskiy_ChemPhysChem_Magnetization_and_Polarization_of_Coupled_Nuclear_Spins_Ensembles_at_High_Magnetic_Fields}
of a given Xe isotope.
Inserting the BCH expansion of Eq.~(\ref{eq:bch_expansion}) into Eq.~(\ref{eq:xe_polarisation})
and evaluating the resulting commutators,
it is shown 
in Appendix A
that the polarization becomes a power series in $A_K \tau$, 
in which every odd-order term vanishes:
\begin{equation}
    \label{eq:xe_polarisation_expansion}
    p_K(\tau) = p_K(0) + \frac{(A_K \tau)^2}{2} f_K + \mathcal{O}\big[(A_K \tau)^4\big].
\end{equation}
The Rela$^2$x~\cite{hil25b, Hilla_2026_Rela2x_Zenodo}
Python software
was used to perform the symbolic calculations and
the necessary scripts are available in the Supplementary Information.
Here, $f_K$ is a factor that depends on the spin quantum number $K$, as described below.
Since the typical
lifetime $\tau$ of a Xe-Rb
collision event is of the order of 10~ps,  
whereas the HFC constant
$A_K$ is of the order of $100$~Mrad/s \cite{ran20, hil26}, the product $A_K \tau$ is
${\cal O}(10^{-3})$.
Thus, Eq.~(\ref{eq:xe_polarisation_expansion}) can be well-approximated
by neglecting the higher-order terms 
$\mathcal{O}\big[(A_K \tau)^4\big]$ of the expansion.
Hence, the SEOP polarization transfer 
to the noble-gas nucleus
during a single collision
event becomes
\begin{equation}
    \label{eq:xe_polarisation_transfer}
    \Delta p_K = p_K(\tau) - p_K(0) \approx \frac{(A_K \tau)^2}{2} f_K.
\end{equation}
The factor $f_K$ is given by
\begin{eqnarray}
f_K &=& \pm \frac{1}{2K(2K + 1)} \times \frac{1}{3} K(K + 1) \dim{\mathcal{H}_K}\nonumber \\
&=& \pm \frac{1}{6} \frac{K + 1}{2K + 1} \dim{\mathcal{H}_K},
\label{eqn-fk}
\end{eqnarray} 
where $\pm \frac{1}{2K(2K + 1)}$ comes
from the definitions of
$\hat{p}_K$ and $\hat{\rho}(0)$,
and $\frac{1}{3} K(K + 1) \dim{\mathcal{H}_K}$ equals the squared
norm ${\rm Tr}\!\left({\hat{K}_z^\dag \hat{K}_z}\right)$ 
(the "Liouville bracket")
of $\hat{K}_z$.
For the electron-nucleus two-spin system, the dimension of the relevant
composite
Hilbert space equals $\dim{\mathcal{H}_K} = 2(2K + 1)$ and, thus,
\begin{equation}
    f_K = \pm \frac{1}{3} (K + 1),
\end{equation}
leading to the final result
\begin{equation}
    \label{eq:xe_polarisation_transfer_final}
    \Delta p_K(\tau) \approx \pm \frac{1}{6} (K + 1) (A_K \tau)^2.
\end{equation}

The ratio of the polarization transfer to $^{129}$Xe and $^{131}$Xe in 
a collision
event with the alkali metal
now becomes
\begin{eqnarray}
    \label{eq:polarisation_transfer_ratio}
    \frac{\Delta p_{129}}{\Delta p_{131}}
    &=& \frac{f_{129}}{f_{131}} \bigg( \frac{A_{129}}{A_{131}} \bigg)^2
    = \frac{K_{129} + 1}{K_{131} + 1} \bigg( \frac{A_{129}}{A_{131}} \bigg)^2 \nonumber \\
    &=& \frac{3}{5} \bigg( \frac{A_{129}}{A_{131}} \bigg)^2.
\end{eqnarray}
The ratio of the HFC constants of the two isotopes is, apart from 
very small corrections caused by the different
nuclear charge and magnetic moment distributions, equal to the ratio of the gyromagnetic
ratios $\gamma_K$ of the isotopes.
Hence, the numerical value of the polarization transfer ratio is given by
\begin{equation}
    \label{eq:polarisation_transfer_ratio_numeric}
    \frac{\Delta p_{129}}{\Delta p_{131}} 
    \approx \frac{3}{5} \bigg( \frac{\gamma_{129}}{\gamma_{131}} \bigg)^2 
    \approx 6.83.
\end{equation}
Due to the leading, quadratic term in 
Eqs.~(\ref{eq:xe_polarisation_expansion}) and (\ref{eq:xe_polarisation_transfer}),
the \emph{square} of 
$\frac{\gamma_{129}}{\gamma_{131}}$
appears in 
the ratio of the polarization 
gain 
instead of 
$\frac{\gamma_{129}}{\gamma_{131}}$
itself, as could naively be anticipated.
This is in line with the theoretical result
for the spin-exchange cross section (see below)
obtained by Herman~\cite{her65}
in the early days of SEOP.

Based on the above theoretical analysis, the
polarization transfer to $^{129}$Xe is expected to
be about 7 times more efficient than that to $^{131}$Xe, 
disregarding the differences in the intrinsic and SEOP cell-wall 
relaxation rates between the two isotopes~\cite{saa15}. 
Since the quadrupolar $^{131}$Xe relaxes much faster than $^{129}$Xe, 
the experimentally observed 
$\frac{\Delta p_{129}}{\Delta p_{131}}$
polarization transfer ratio can be expected
to be always \textit{larger} than the theoretical value of 6.83.

It is interesting to note that if the gyromagnetic ratios
of the two isotopes were the same, polarization transfer
to the spin-$\frac{3}{2}$ 
$^{131}$Xe would be more efficient than that to
the spin-$\frac{1}{2}$ $^{129}$Xe
by a factor of $\frac{5}{3} \approx 1.67$.
This arises from the larger norm of the $\hat{K}_z$ operator
for 
$^{131}$Xe than for 
$^{129}$Xe:
the same increase in the amplitude of 
the polarization level operator
$\hat{p}_K$ in $\hat{\rho}$ causes a larger
increase in the $z$ polarization level of $^{131}$Xe 
than for 
$^{129}$Xe.
Another interesting remark is that twice the ratio of the HFC constants,
$2\frac{A_{129}}{A_{131}} = 2\frac{\gamma_{129}}{\gamma_{131}} \approx 6.75$,
happens to be very close to the actual 
polarization transfer
ratio of about 6.83
obtained above. Again, one might naively expect that 
the two ratios are connected,
but this is purely coincidental.

We note that in SEOP literature one often 
refers to the spin-exchange
cross section, which is 
related to $\hat{K}_z$ instead of the polarization level $\hat{p}_K$.
Had we used $\hat{K}_z$ in Eq.~(\ref{eq:xe_polarisation}) 
to 
define the polarization of Xe, the polarization
transfer ratio would have become $\frac{K_{129}}{K_{131}} \frac{\Delta p_{129}}{\Delta p_{131}}$,
as reported to be the ratio of the spin-exchange cross sections in, {\em e.g.}, Ref.~\cite{mol23}
where the results 
obtained by Herman~\cite{her65} are used.
The corresponding ratio 
between the two isotopes would then be
$\frac{K_{129}}{K_{131}} \times 6.83 = \frac{1}{3} \times 6.83 = 2.28$.
It is therefore important to pay attention to
what exactly is meant
by ``polarization transfer" in the case of SEOP.
In the present work, we always refer to the polarization level of Xe
as defined by Eq.~(\ref{eq:xe_polarisation}), which is
in line with how polarization is perceived in the magnetic resonance
community.

\section{Simulations}

\subsection{Molecular dynamics}

Classical 
MD simulations were performed to
generate the Rb-Xe interaction events that mediate the polarization transfer. 
All simulations were performed using the LAMMPS \cite{Thompson2022} 
software
with a simulation cell containing 
a simplified SEOP gas mixture of
one Rb atom and 2196 Xe atoms.
The Rb-Xe and Xe-Xe interactions were described by the 
diatomic
\textit{ab initio} 
potential energy curves of Hanni \textit{et al.}~\cite{han17} and Hellmann, J\"ager, and Bich~\cite{Hellmann2017}, respectively. 
The simulations were carried out at a temperature $T = 300$ K and pressure $P = 2.37$ atm, with a time step of $\Delta t = 1$~fs. For more details we refer to Appendix B.

The MD trajectories were post-processed to identify and classify the Rb-Xe interaction events. As before~\cite{hil26,ran20}, an interaction event was defined as any instance in which a Xe atom approaches within a cutoff radius of $r_c = 9$ \AA\ of the Rb 
atom.
Based on the subsequent behavior of the Rb-Xe pair,
interaction events were classified into two types: 
short-lived scattering collisions, 
in which Rb and Xe approach and promptly rebound, and vdW complexes, where they form a relatively long-lived
metastable bound state. 
More details on the classification algorithm are provided in Ref.~\cite{hil27}.
Applying the above criteria,
a total of 111,227 scattering and 765 vdW events were identified in the simulation.

\subsection{Spin dynamics}


Details of the multiscale modeling procedure for polarisation transfer in SEOP, 
combining the MD simulations with
QC spin Hamiltonian parameter calculations and 
SD simulations, are reported in Ref.~\cite{hil26}.
Briefly, each Rb-Xe collision event $\epsilon$ extracted from the MD trajectory
defines a time series of Rb-Xe interatomic
distances $\{R^\epsilon(n\Delta \tau)\} \equiv \{R^\epsilon_n\}$,
where $n$ is the index of the MD
frame and $\Delta \tau = 50$ fs is
the time interval between the frames. 
The instantaneous
spin Hamiltonian $\hat{H}\big[R(t)\big] \equiv \hat{H}(t)$ of
the Rb-Xe two-atom 
system is completely determined by its only degree of freedom $R(t)$.
The full spin Hamiltonian used in the present SD simulations is given in Appendix~C.

The time series of the
interatomic distances $\{R^\epsilon_n\}$ defines
a corresponding time series of spin Hamiltonians 
$\{\hat{H}^\epsilon_n\}$,
that, in turn, determines the Hilbert-space
time evolution of the spin density operator $\hat{\rho}^\epsilon(t)$ 
for each event $\epsilon$ as
\begin{equation}
    \hat{\rho}^\epsilon \big[(n + 1) \Delta \tau\big] = 
    e^{-i \hat{H}^\epsilon_n \Delta \tau} \hat{\rho}^\epsilon(n\Delta \tau) e^{i \hat{H}^\epsilon_n \Delta \tau},
\end{equation}
from which the time evolution 
of the 
polarization level of Xe, $p_K^\epsilon(t)$,
can be extracted with Eq.~(\ref{eq:xe_polarisation}).
The ensemble average
\begin{equation}
    \bigg\langle \frac{\Delta p_{129}}{\Delta p_{131}} \bigg\rangle_{\{\epsilon\}} =
    \frac{1}{N} \sum_{\epsilon=1}^{N} \frac{\Delta p_{129}^\epsilon(\tau_\epsilon)}{\Delta p_{131}^\epsilon(\tau_\epsilon)}
\label{eq:ensemble_average}
\end{equation}
over the collision events yields the average polarization transfer ratio per event, 
to be compared with the theoretical prediction of Eq.~(\ref{eq:polarisation_transfer_ratio_numeric}).
The SD simulations were always
initialised with the spin state of the Rb atom being
fully polarised
as a result of the optical pumping process, and the Xe nucleus in a
thermal equilibrium state. 
A magnetic field strength of $B_0 =1$~mT, characteristic of 
SEOP experiments, was used.
All SD simulations were performed
using an in-house developed Python software \cite{hil26},
which is
available upon request.

\section{Simulation results and discussion}

\begin{figure}[h]
    \centering
    \includegraphics[width=0.95\columnwidth]{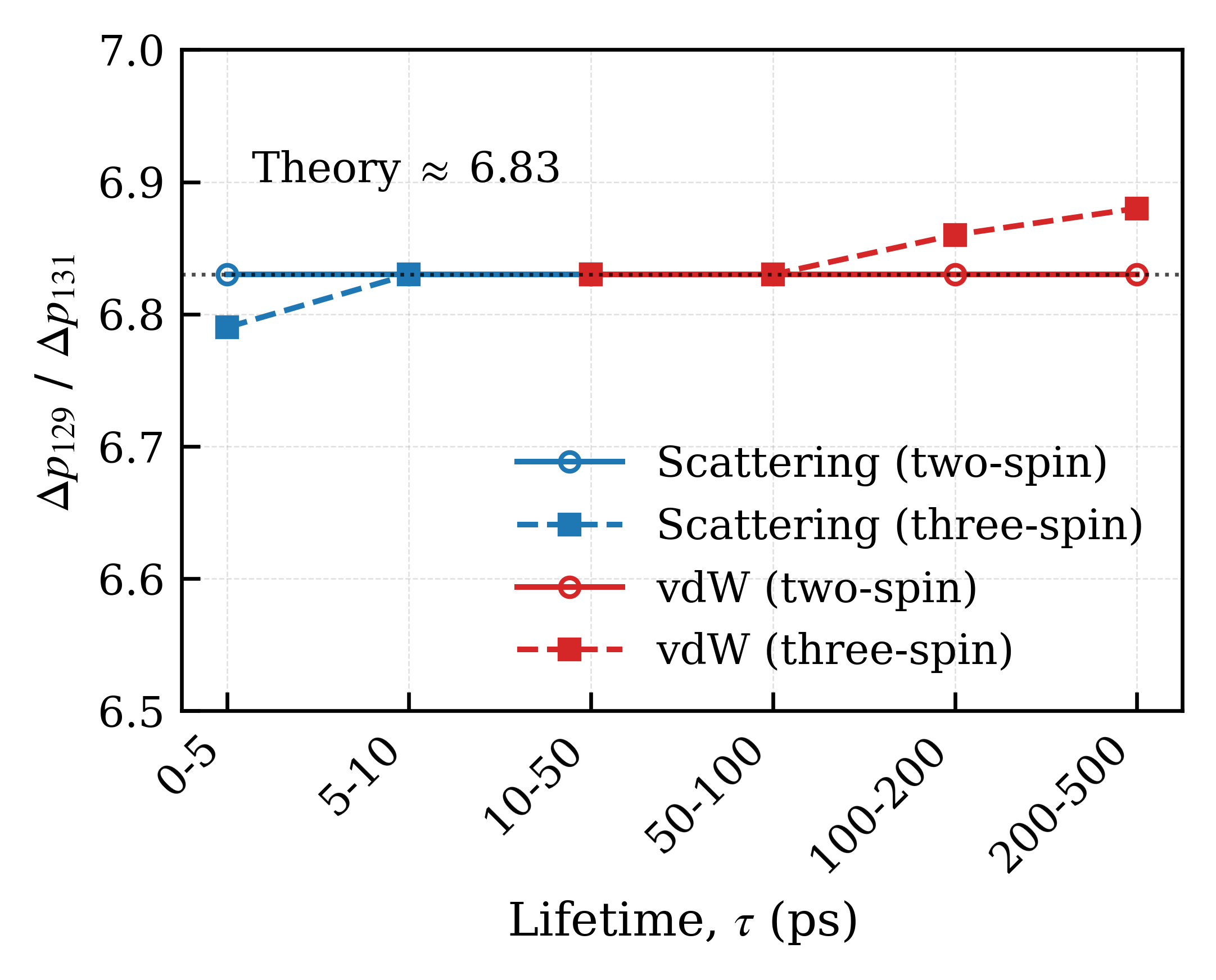}
     \caption {
    Simulated
    polarization transfer ratios from Table~\ref{tab:ratios} for scattering and
    vdW events, in the two- and three-spin models
    (with and without the $^{87}$Rb nuclear spin).
    The dotted line indicates the theoretical prediction of 6.83.
     }
    \label{fig:ratio}
\end{figure}

\begin{table}[t]
\setlength{\tabcolsep}{5pt}
\centering
\caption{Simulated per-event average polarization transfer ratio 
$\big\langle \Delta p_{129}/\Delta p_{131} \big\rangle_{\{\epsilon\}}$ 
[Eq.~(\ref{eq:ensemble_average})], 
across collision-lifetime categories for scattering and vdW events. $N$ is the number 
of events per category. Results are given for both the two-spin 
(no Rb nucleus) and three-spin (with Rb nucleus) models. The 
statistical
uncertainty (SEM) of each ratio is below $0.02$.
   }
\vspace{0.5cm}
\label{tab:ratios}
\begin{tabular}{l r c c}
\hline\hline
\makecell{Lifetime\\(ps)}
&
\makecell{$N$}
&
\makecell{$\big\langle \Delta p_{129}/\Delta p_{131} \big\rangle_{\{\epsilon\}}$\\[-1pt]\scriptsize(two-spin)}
&
\makecell{$\big\langle \Delta p_{129}/\Delta p_{131} \big\rangle_{\{\epsilon\}}$\\[-1pt]\scriptsize(three-spin)}
\\
\hline

\multicolumn{4}{l}{\textit{Scattering}} \\

0--5
& 98074
& 6.83
& 6.79 \\

5--10
& 11887
& 6.83
& 6.83 \\

10--50
& 1266
& 6.83
& 6.83 \\

\hline
\multicolumn{4}{l}{\textit{vdW}} \\

10--50
& 434
& 6.83
& 6.83 \\

50--100
& 240
& 6.83
& 6.83 \\

100--200
& 82
& 6.83
& 6.86 \\

200--500
& 9
& 6.83
& 6.88 \\

\hline\hline
\end{tabular}%
\end{table}
Table~\ref{tab:ratios} lists the simulated
average polarization transfer ratios
according to Eq.~(\ref{eq:ensemble_average})
in different collision event
lifetime categories
and for the two different kinds of spin-system models.
We observe that 
the 
polarization transfer to ${}^{129}$Xe is
significantly higher than that for $^{131}$Xe. This is consistent with previous experiments~\cite{vol80,kwo81,sha05,but94,stu11,Lu2024}
and 
the theoretical analysis in Section~\ref{sec-theory}.
As illustrated in Fig.~\ref{fig:ratio}, 
regardless of whether the Rb nuclear spin
is included
(three-spin model) or not (two-spin model), 
the polarization transfer ratio
remains remarkably close to the
theoretically predicted value of 6.83.
The prediction is, thus, verified by the present computational experiment.

Some deviation from the theoretical value is observed for the very short-lived scattering events, 
when the Rb nuclear spin is included. 
We believe that this arises simply
because the magnitudes of $\Delta p_{K}$ for the shortest-lived events
are extremely
small (order of $10^{-13}$ or less), and the ratio of two such small numbers is sensitive
to numerical inaccuracies in the simulations, which have not been tuned for such extreme precision.
We thus do not consider this deviation to have any physical origin.

There is also a slight deviation for the long-lived vdW events when the Rb nucleus is included,
which does have a physical origin:
The internal spin dynamics of the Rb atom
is dictated by intra- and inter-branch hyperfine transitions, which are characterized by different Rabi oscillation frequencies~\cite{hil26}. Therein, the 
spin system
oscillates between the initial and final SEOP states, and this provides an envelope over which the much faster dynamics of the 
vdW bond between the Rb and Xe atoms is superimposed.
Significant polarization transfer to Xe occurs when the 
Rb-Xe interatomic distance is near the minimum value of the vdW bond length.
The Rabi frequency depends on the transition energy $\hbar\omega_{fi}$ and Xe HFC $A_K$ as~\cite{hil26} 
\begin{equation}
\omega=\left(\omega_{fi}^2+a_{fi}A_K^2\right)^{1/2},
\end{equation}
where $a_{fi}$ depends on the matrix element of the 
flip-flop operator between the initial and final states, $i$ and $f$. Each SEOP transition corresponds to its own oscillation frequency, and taking into account the limited lifetime of the vdW complexes, the most relevant transitions occur between the upper and lower hyperfine branches of the Rb spectrum. 
$A_{129}$ is larger than $A_{131}$ by a factor
of $|\frac{\gamma_{129}}{\gamma_{131}}| \approx 3.4$, and
hence the relevant inter-branch transition has a shorter period for $^{129}$Xe.
Consequently, the envelope for polarization transfer increases faster for $^{129}$Xe than for $^{131}$Xe, which leads to $\Delta p_{129}/\Delta p_{131}$ gaining, on average, values exceeding the theoretical result of 6.83, in the longest events and within the three-spin model that features such intra-branch transitions. Appreciable differences from the initial polarization state of Rb due to these oscillations are only seen after after a few tens of ps.
    Hence, this only affects the vdW events that have a sufficiently long lifetime for the Rb spin dynamics to manifest. The theoretical
    model introduced in the present paper does not account for this, which is the reason for the slight deviation in the case of the three-spin model for the longest-lived events.

Early experiments have placed the ratio of the spin-exchange cross sections $\sigma_{129}/\sigma_{131}$,
which translates to the absolute polarization transfer ratio $\Delta K_z(129)/\Delta K_z(131)$, 
at 3.56 (Ref.~\cite{kwo81}) and 16 (Ref.~\cite{vol80}), based on relaxation data and assuming binary 
(scattering)
collisions to dominate.
More recently, Molway {\em et al.}~\cite{mol23} reached maximally $(7.6\pm 1.5)$\% polarization of $^{131}$Xe, a larger value than the pioneering result of 2.2\% by Stupic {\em et al.}~\cite{stu11}. 
In Ref.~\cite{mol23}, using the results
from Herman \cite{her65}, the ratio of 2.28 for 
$\Delta K_z(129)/\Delta K_z(131)$ was used as a reference, as discussed in Section~\ref{sec-theory}.
Results for the 
$\sigma_{129}/\sigma_{131}$ at 2.5\ldots 5 in Rb-Xe mixtures~\cite{mol23} and 1.7\ldots 3 (Ref.~\cite{mol23}) or 1.3\ldots 3.6 (Ref.~\cite{sha05}) in Cs-Xe mixtures have been obtained. 
Finally, Lu {\em et al.}~\cite{Lu2024} obtained polarization of 17.6\% and 1.96\% for $^{129}$Xe and $^{131}$Xe, respectively, in two separate SEOP cells and conditions. This would correspond to a ratio of $\Delta p_{129}/\Delta p_{131}=8.98$, albeit this number is less reliable due to the different experimental conditions.
Overall, all of these results are in agreement with the presently obtained theoretical lower limit of $\Delta K_z(129)/\Delta K_z(131) = 2.28$ and $\Delta p_{129}/\Delta p_{131} = 6.83$, with the experimental results being larger due to the faster relaxation of the quadrupolar $^{131}$Xe isotope.

\section{Conclusions}


We have presented a simple theoretical model for the SEOP polarization transfer during gas-phase collisions between the alkali-metal and noble-gas atoms. 
The model is based on the leading-order non-vanishing (quadratic)
term in the Baker-Campbell-Hausdorff expansion of the time evolution of the spin density operator $\hat{\rho}(t)$,
which is valid due to the short-lived nature of the collision events. 
For a noble-gas atom with nuclear spin $K$,
the polarization transfer $\Delta p_K(\tau) \approx \pm \frac{1}{6} (K + 1) (A_K \tau)^2$
is predicted, where $\tau$ is the duration of the collision event and $A_K$ the average
hyperfine coupling between the alkali-metal electron and noble-gas nucleus.

The model assumes a fully spin polarized alkali-metal electron and neglects spin relaxation, 
and therefore gives the upper limit of the polarization transfer to the noble-gas nucleus.
We applied the model to study the degree of polarization gained
in SEOP by the spin-1/2 $^{129}$Xe and spin-3/2 $^{131}$Xe isotopes, 
which are both used in many magnetic resonance applications.
The polarization transfer ratio 
$\Delta p_{129}/\Delta p_{131} = \frac{3}{5}\left(\frac{\gamma_{129}}{\gamma_{131}}\right)^2 \approx 6.83$ 
is predicted, where $\gamma_K$ is the gyromagnetic ratio of the isotope with nuclear spin $K$,
indicating that in comparable conditions, the polarization transfer to $^{131}$Xe is significantly
weaker than that to $^{129}$Xe.
This ratio was verified by numerical multiscale simulations combining quantum-chemically parameterized
spin dynamics of Rb--Xe collision events extracted from molecular dynamics simulations,
and is in good agreement with previous experimental results.
Slight deviations from the predicted ratio for 
very long-lived collision events
were rationalized 
by the internal spin dynamics of the Rb atom.

In experimental work, the faster relaxation of the quadrupolar $^{131}$Xe isotope renders the observable
$\Delta p_{129}/\Delta p_{131}$ ratio even larger than the presently predicted theoretical value of 6.83.
The theoretical value can therefore be considered as a lower bound for the actual polarization transfer ratio in experiments.

The present work provides a simple theoretical framework for understanding the polarization transfer in SEOP, 
particularly for spin $I > 1/2$ noble-gas isotopes, and thereby contributes toward rational design of
improved SEOP hyperpolarization setups.

\begin{acknowledgments}

Financial support has been obtained from the  University of Oulu (Kvantum Institute) and the Research Council of Finland (project 361326). 
PH is grateful for
funding from 
University of Oulu graduate school (UniOGS), the Finnish Cultural Foundation and the Emil Aaltonen Foundation.
Computational resources due to CSC---IT Center for Science (Espoo, Finland) were used.

\end{acknowledgments}

\appendix

\section{Evaluation of the commutators}

The following discussion treats the
symbolic evaluation of the commutators in Eq.~(\ref{eq:bch_expansion})
using the Rela$^2$x~\cite{hil25b, Hilla_2026_Rela2x_Zenodo} and SymPy~\cite{Meurer_etal_PeerJComputSci_2017_SymPy_symbolic_computing_in_Python} Python libraries.
The associated Jupyter Notebook is given in the Supplemental Material. 

For the electron-$^{129}$Xe two-spin system,
the first-order commutator 
$\hat{C}_1 \equiv [\hat{H}_{\rm HFC}^{(K)}, \hat{\rho}(0)]$
in Eq.~(\ref{eq:bch_expansion}),
expressed in the direct product basis $\{\hat{b}_i\}$
of 
(non-normalized)
irreducible spherical tensor (IST)
operators of the form
$\hat{T}_{l, q}^{(S)} \hat{T}_{l', q'}^{(K)}$,
which span the entire operator (Liouville) space of the spin system,
becomes
\begin{equation}
\hat{C}_1 = \pm \frac{A_K}{2K + 1}
\bigg( \hat{T}_{1, 1}^{(S)} \hat{T}_{1, -1}^{(K)} - \hat{T}_{1, -1}^{(S)} \hat{T}_{1, 1}^{(K)} \bigg).
\end{equation}
The polarization level operator 
can be expressed in terms of the IST operators
as $\hat{p}_K = \hat{K}_z/K = \hat{T}_{1, 0}^{(K)}/K$, 
where $\hat{T}_{1, 0}^{(K)}$ 
is one of the operators of the
basis $\{\hat{b}_i\}$.
The term proportional to
${\rm Tr}\!\left({\hat{p}_K \hat{C}_1}\right) = {\rm Tr}\!\left({\hat{p}_K^\dagger \hat{C}_1}\right)$
($\hat{p}_K$ is Hermitian)
in Eq.~(\ref{eq:xe_polarisation})
vanishes, because
the IST
basis operators are orthogonal to each other,
such that ${\rm Tr}\!\left({\hat{b}_i^\dagger \hat{b}_j}\right) \propto \delta_{ij}$.
In contrast, the second-order commutator $\hat{C}_2 \equiv [\hat{H}_{\rm HFC}^{(K)}, \hat{C}_1]$ 
becomes
\begin{equation}
\hat{C}_2 = \pm \frac{A_K^2}{2(2K + 1)} \bigg( \hat{T}_{1, 0}^{(S)} - \hat{T}_{1, 0}^{(K)} \bigg),
\end{equation}
which yields
\begin{eqnarray}
\label{eq:trace_C2}
{\rm Tr}\!\left({\hat{p}_K^\dagger \hat{C}_2}\right) &
= &\pm \frac{A_K^2}{2K(2K + 1)} {\rm Tr}\!\left[{\hat{T}_{1, 0}^{\dagger (K)} \hat{T}_{1, 0}^{(K)}}\right] \nonumber \\
& = &\pm \frac{A_K^2}{2K(2K + 1)} {\rm Tr}\!\left({\hat{K}_z^\dagger \hat{K}_z}\right).
\end{eqnarray}

For the electron-$^{131}$Xe two-spin system, we obtain the same expression
for $\hat{C}_1$ as for $^{129}$Xe
above, but the second-order commutator becomes, this time, 
\begin{eqnarray}
\hat{C}_2 = \pm \frac{A_K^2}{6(2K + 1)} 
\left\{ 3\left[
5\hat{T}_{1, 0}^{(S)} - \hat{T}_{1, 0}^{(K)} 
+ \sqrt{2} \hat{T}_{1, -1}^{(S)} \hat{T}_{2, 1}^{(K)} \right. \right.
\nonumber \\
\left. \left. \mbox{} + \sqrt{2} \hat{T}_{1, 1}^{(S)} \hat{T}_{2, -1}^{(K)} 
\right] 
 - 2\sqrt{6} \hat{T}_{1, 0}^{(S)} \hat{T}_{2, 0}^{(K)} 
\right\}.
\end{eqnarray}
However, when taking the trace with $\hat{p}_K$, 
the same result as for $^{129}$Xe is obtained.

To evaluate the squared norm ${\rm Tr}\!\left({\hat{K}_z^\dagger \hat{K}_z}\right)$
appearing in Eq.~(\ref{eq:trace_C2}),
we note that $\hat{K}_z = \hat{1}_S \otimes \hat{K}_{z, K}$,
where the first and second  operator 
act on the Hilbert spaces of
the electron and the Xe nuclear spins, respectively. 
Using the properties of the trace
and the hermiticity of 
$\hat{K}_{z, K}$, we get
\begin{eqnarray}
{\rm Tr}\!\left({\hat{K}_z^\dagger \hat{K}_z}\right) 
& = & {\rm Tr}\!\left[{(\hat{1}_S \otimes \hat{K}_{z, K})^\dagger (\hat{1}_S \otimes \hat{K}_{z, K})}\right] \nonumber\\
& = & {\rm Tr}\!\left({\hat{1}_S^\dagger \hat{1}_S}\right) {\rm Tr}\!\left({\hat{K}_{z, K}^\dagger \hat{K}_{z, K}}\right) \nonumber \\
& = & (2S + 1) {\rm Tr}\!\left({\hat{K}_{z, K}^2}\right).
\end{eqnarray}
We then
note that the trace of a matrix is basis
independent, so it is most convenient
to use the Zeeman basis of the Xe spin, 
in which $\hat{K}_{z, K}$ is diagonal
with eigenvalues $-K, -K + 1, \ldots, K$. This yields
\begin{equation}
{\rm Tr}\!\left({\hat{K}_{z, K}^2}\right) = \sum_{m_K = -K}^K m_K^2 =
\frac{1}{3} K(K + 1)(2K + 1).
\end{equation}
Noting that $(2S + 1)(2K + 1) = \dim(\mathcal{H})$, where $\mathcal{H}$ is
the Hilbert space of the composite electron-Xe spin system,
we get
\begin{equation}
{\rm Tr}\!\left({\hat{K}_z^\dagger \hat{K}_z} \right)
= \frac{1}{3} K(K + 1) \dim(\mathcal{H}).
\end{equation}
This yields
\begin{equation}
{\rm Tr}\!\left({\hat{p}_K \hat{C}_2}\right) = A_K^2 f_K,
\end{equation}
where
\begin{eqnarray}
f_K &=& \pm \frac{1}{2K(2K + 1)} \times \frac{1}{3} K(K + 1) \dim{\mathcal{H}_K} \nonumber \\
&=& \pm \frac{1}{6} \frac{K + 1}{2K + 1} \dim{\mathcal{H}_K},
\end{eqnarray}
as given in the main text.

The third-order commutator $\hat{C}_3 \equiv [\hat{H}_{\rm HFC}^{(K)}, \hat{C}_2]$
is proportional to the first-order commutator $\hat{C}_1$
and, hence, it is easily seen that every odd-order commutator vanishes when taking the trace with $\hat{p}_K$.

\section{Molecular dynamics details}

\begin{figure}[h]
     \centering
     \includegraphics[width=0.45\textwidth]{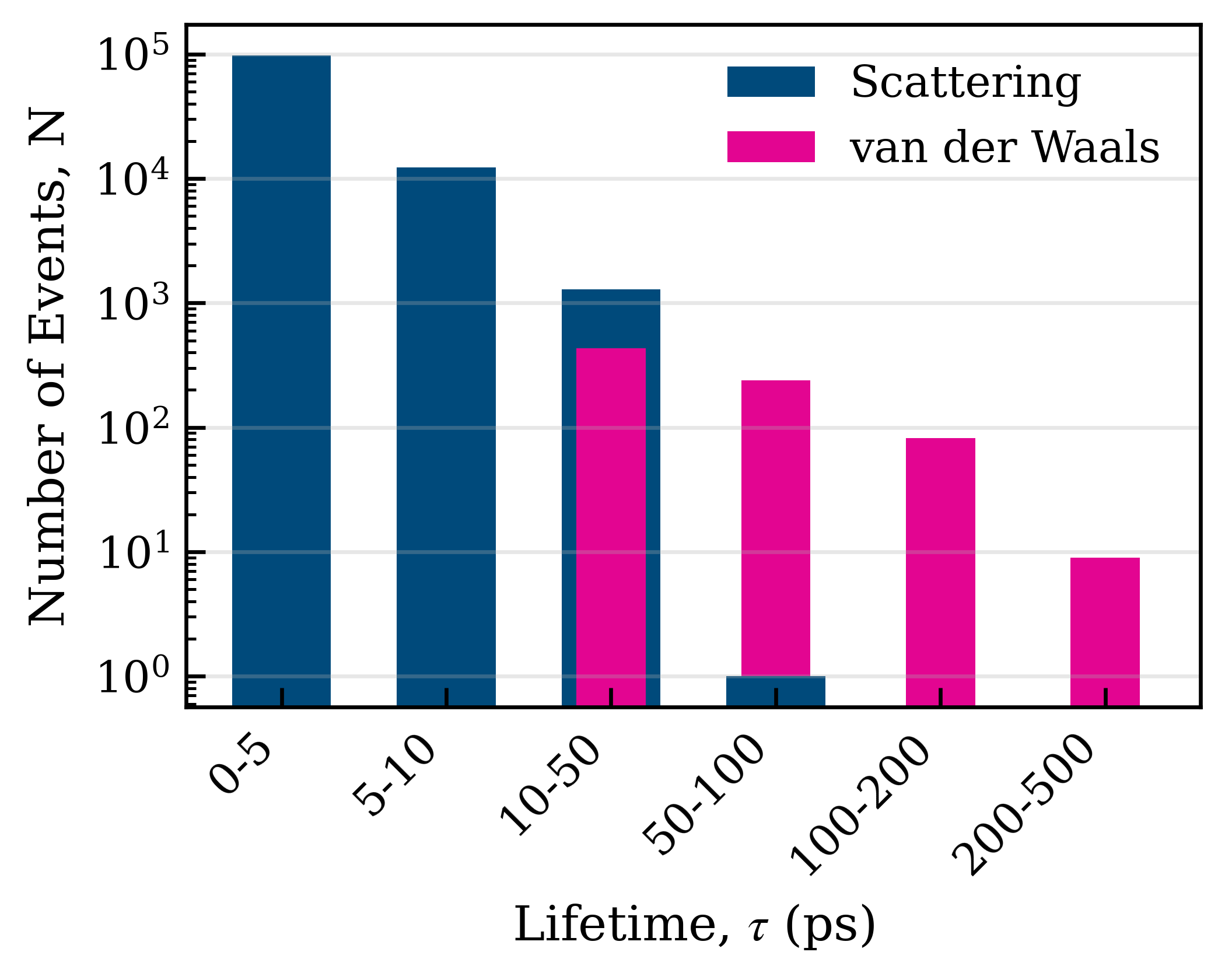}
     \caption{Distribution of Rb-Xe collision
    event lifetimes
    for the scattering and vdW event categories.
     }
     \label{fig:duration_by_category}
\end{figure}

The initial atomic configurations were generated using PACKMOL~\cite{Martnez2009}. The system was equilibrated through a three-stage protocol: a 1-ns 
isothermal-isobaric $NPT$
run with a Nos\'e-Hoover thermostat \cite{nose1984unified, hoover1985canonical} and barostat \cite{martyna1994constant, Shinoda2004},
using
temperature and pressure damping parameters of $\tau_T = 100$~fs and $\tau_P = 100$~fs, respectively, followed by a 1-ns 
canonical $NVT$
run, and finally a 1-ns microcanonical 
$NVE$ run.
The $NPT$ stage yielded an equilibrium cubic box length of $L = 332.43$ \AA, which was retained for all subsequent simulation stages. The equilibrated configuration was then propagated for 
2~$\mu$s in the $NVE$ ensemble, with atomic positions and velocities recorded every 50~fs, producing a total 
of $4 \times 10^7$ frames for analysis.
Lifetime distributions of the 
scattering and vdW events are shown in Fig.~\ref{fig:duration_by_category}. 
The average duration of a scattering event is $3.217\pm0.006$~ps and that of a vdW event is $56.1\pm 1.5$~ps. 

\section{Spin Hamiltonian}

The three-spin Hamiltonian 
used in the present spin dynamics simulations
is given by
\begin{eqnarray}
    \hat{H}(t) & = &\mu_{\rm B} \hat{\mathbf{S}} \cdot \mathbf{g}(t) \cdot \mathbf{B} 
    + \mu_N g_{\text{Rb}} \hat{\mathbf{I}} \cdot \mathbf{B} 
    + \mu_N g_{\text{Xe}} \hat{\mathbf{K}} \cdot \mathbf{B} \nonumber \\
    && \mbox{} + h \, \hat{\mathbf{S}} \cdot \mathbf{A}_{\text{Rb}}(t) \cdot \hat{\mathbf{I}}
    + h \, \hat{\mathbf{S}} \cdot \mathbf{A}_{\text{Xe}}(t) \cdot \hat{\mathbf{K}} \nonumber \\
    && \mbox{} + \hat{\mathbf{S}} \cdot \bm{\epsilon}(t) \cdot \mathbf{M} (t) + h \hat{\mathbf{K}} \cdot \mathbf{Q}_{\text{Xe}}(t) \cdot \hat{\mathbf{K}},
\end{eqnarray}
where $\mu_{\rm B}$ and $\mu_N$ are the Bohr and nuclear magnetons, respectively,
$g_{\text{Rb}}$ and $g_{\text{Xe}}$ are the nuclear $g$-factors,
$\hat{\mathbf{S}}$, $\hat{\mathbf{I}}$, and $\hat{\mathbf{K}}$ are the dimensionless
electron, Rb nuclear, and Xe nuclear spin operators, respectively,
$\mathbf{g}(t)$ is the electron $g$-tensor, $\mathbf{B}$ is the external magnetic field,
$\mathbf{A}_{\text{Rb}}(t)$ and $\mathbf{A}_{\text{Xe}}(t)$ are the HFC tensors between
the electron and the Rb and Xe nuclei,
$\bm{\epsilon}(t)$ is the electron
spin rotation tensor, $\mathbf{M}(t)$ is the mechanical
angular momentum of the colliding Rb-Xe pair,
and 
$\mathbf{Q}_{\text{Xe}}(t)$ 
is the nuclear 
quadrupole coupling tensor of 
$^{131}$Xe.
For details of the QC calculations of the spin interaction tensors, 
see Ref.~\cite{hil26}. The $\mathbf{A}_{131}$ tensor was obtained from $\mathbf{A}_{129}$ by multiplying with $\gamma_{131}/\gamma_{129}$. The $\mathbf{Q}_{131}$ tensor was obtained using the same methodology as Rb $\mathbf{Q}$ in Ref.~\cite{hil26} and the results are listed in Table~\ref{tbl-QXe}. In the two-spin model comprising only the unpaired electron and the Xe nucleus, the nuclear interactions of Rb (terms involving $g_{\rm Rb}$ and $\mathbf{A}_{\rm Rb}$)
were omitted.
\begin{table}[ht]
\setlength{\tabcolsep}{18pt}
\caption{Calculated 
$^{131}$Xe 
quadrupole coupling tensor $\bm{Q}$ (in MHz). Parallel ($Q_{\parallel}$) and perpendicular ($Q_\perp$) components, referring to the internuclear axis of the Rb-Xe atom pair.} 
\label{tbl-QXe}
\begin{center}
\begin{tabular}{lcc}\hline\hline
$R$ (\AA) & 
$Q_\parallel\!({^{131}}{\rm Xe})$ & $Q_\perp\!({^{131}}{\rm Xe})$ 
\\ \hline
3.0 & $0.7661$ & $-0.3830$  \\
3.2 & $-0.7451$ & $0.3725$  \\
3.4 & $-1.5121$ & $0.7558$  \\
3.6 & $-1.7661$ & $0.8830$  \\
3.8 & $-1.7094$ & $0.8545$  \\
4.0 & $-1.5045$ & $0.7522$  \\
4.2 & $-1.2527$ & $0.6263$  \\
4.4 & $-1.0067$ & $0.5034$  \\
4.6 & $-0.7880$ & $0.3940$  \\
4.8 & $-0.6125$ & $0.3063$  \\
5.0 & $-0.4683$ & $0.2342$  \\
5.2 & $-0.3589$ & $0.1795$  \\
5.4 & $-0.2705$ & $0.1350$  \\
5.6 & $-0.2040$ & $0.1018$  \\
5.8 & $-0.1518$ & $0.0759$  \\
6.0 & $-0.1116$ & $0.0558$  \\
6.2 & $-0.0835$ & $0.0417$  \\
6.4 & $-0.0612$ & $0.0308$  \\
6.6 & $-0.0478$ & $0.0237$  \\
6.8 & $-0.0344$ & $0.0174$  \\
7.0 & $-0.0290$ & $0.0147$  \\
7.4 & $-0.0214$ & $0.0107$  \\
7.8 & $-0.0156$ & $0.0078$  \\
8.2 & $-0.0147$ & $0.0074$  \\
8.6 & $-0.0121$ & $0.0060$  \\
9.0 & $-0.0125$ & $0.0063$  \\
9.4 & $-0.0085$ & $0.0040$  \\
9.8 & $-0.0085$ & $0.0040$  \\
10.2 & $-0.0063$ & $0.0031$  \\
10.6 & $-0.0054$ & $0.0027$  \\
11.0 & $-0.0045$ & $0.0022$  \\
\hline\hline\end{tabular}\end{center}
\end{table}

\bibliography{references}

\end{document}